\documentclass[11pt,letterpaper]{article}
\usepackage[table,svgnames]{xcolor}
\usepackage{graphicx}
\usepackage{amsmath,amssymb}
\usepackage{subfigure}
\usepackage{url}
\usepackage{verbatim}
\usepackage{times}
\usepackage{wrapfig}
\usepackage{ifpdf}
\usepackage[top=0.8in,bottom=0.85in,left=0.8in,right=0.8in]{geometry}
\newcommand{\arxiv}{arXiv}
\newcommand{\latex}{\LaTeX}
\renewcommand{\paragraph}[1]{\medskip\noindent{\bf #1}}
\newcommand{\qed}{\ensuremath{\square}}
\newcommand{\eat}[1]{}
\ifpdf\fi

\date{}

\title{Scienceography: the study of how science is written}
\author{
Graham Cormode \\
{AT\&T Labs--Research}
\and
S. Muthukrishnan \\
{Rutgers University}
\and
Jinyun Yan \\
{Rutgers University}
}

\begin{document}
\raggedbottom
\maketitle

\begin{abstract}
Scientific literature has itself been the subject of much scientific
study, for a variety of reasons: understanding how results are
communicated, how ideas spread, and assessing the influence of areas
or individuals.  
However, most prior work has focused on extracting and analyzing
citation and stylistic patterns. 
In this work, we introduce the notion of `scienceography', which
focuses on the {\em writing} of science. 
We provide a first large scale study using data derived from
the \arxiv\ e-print repository. 
Crucially, our data includes the ``source code'' of scientific papers---the \latex\ source---which enables us to study features not present in the ``final product'',
such as the tools used and private comments between authors. 
Our study identifies broad patterns and trends in two example areas---computer science and mathematics---as well as highlighting key
differences in the way that science is written in these fields. 
Finally, we outline future directions to extend the new
topic of scienceography. 
\end{abstract}

\section{Introduction}

\label{sec:intro}
Many people seek to understand the progress of science by studying
aspects of the process by which new scientific knowledge is created.
Anecdotes and mythology abound about the process of discovery of
scientific principles and design of new methodologies.
Consider, for example, the
narratives surrounding Newton's Theory of Gravity, 
or Archimedes' invention of a way to measure the volume of solid objects. 
Likewise, there is much study of how scientific knowledge is
propagated through the scientific literature. 
The area of bibliometrics concerns itself with measuring properties of
the research corpus, in particular, the citation patterns among
texts \cite{Atkins:Cronin:00,Klink:Reuther:Weber:Walter:Ley:06,DeBellis:09,Moed:11}. 
This leads to measures of importance, based on notions such as
the citation count of a paper, the impact factor of a journal and the
h-index of an author~\cite{hindex}. 
The specific application of measurement to scientific impact is known as scientometrics, and is chiefly concerned with analyzing and proposing bibliometric measures. 
There is also study of social aspects associated with
scientific
research, such as the ``sociology of scientific knowledge'',
including social structures and processes of scientific activity, as well as policy aspects of Science research~\cite{polurl}. 

Yet, between initial discovery and the dissemination of papers, 
there has been little focus 
on the process of {\em describing} scientific results, in the form of
papers. 
While bibliometrics and sociology of sciences concern themselves with the after-effects of this
work, we have relatively little insight into how the writing
of science is performed. 
In part, this is due to the lack of visibility into this process and the intermediate steps. 
In a few cases, the notes and working papers of notable scientists have
been made available, and these have been studied on an individual
basis. 
But there has been no large scale study, in contrast to the analysis
of citation networks containing thousands to millions of citations.
Recently, there have been efforts to capture trends and influence in science, based on using both citation relations and extracted text from document collections~\cite{Gerrish:Blei:10,Goth:12}. 
The area of quantitative data analysis also applies to track i.e. common words and bursts of interest in particular topics.  
Our aim is to go deeper, and learn about structures within science writing beyond the ``bag of words'' in each paper. 

In this paper, we identify the study of this part of the scientific
method as a topic of interest, which we call  {\em scienceography} 
(meaning ``the writing of science'').
We identify a source of data that allows us to begin to measure
scienceographic properties. 
Using this data, we are able to quantify certain key properties of science writing, its processes, and
how they vary between related areas, and across time. 

Our work proceeds as follows.
In Section~\ref{sec:data}, we describe our data collection from the 
\arxiv, a large collection of scientific reports.
A vital property of the \arxiv\ is that many papers are available in 
\latex\ format, a mark-up language that enables scienceographic
study. 
Section~\ref{sec:analysis} gives our initial analysis on two related
areas, mathematics and computer science, and we compare features of
the writing process. 
These include the use of comments to keep notes,
communicate to co-authors, and adjust text;
the use of additional tools such as macros and packages to facilitate the writing process; 
and the use of figures and theorems to illustrate the authors' intent.%communicate the authors' intent. 
Finally, we conclude by the general aspects of scienceography, and leave directions for further study. 

\section{Data collection}
\label{sec:data}
\paragraph{The \arxiv.}
Our study of scienceography was performed over the \arxiv\ technical
report service. 
The arXiv is an open-access web-based e-print repository that covers
many scientific fields, including physics, mathematics,
nonlinear sciences, computer science, quantitative biology,
quantitative finance and statistics\footnote{\url{http://arxiv.org}}. 
Across all areas, over 700,000 documents have been made available via
the service. 
The service began in 1991, and is 
 primarily maintained and operated by the Cornell University Library. 
After registration, users may upload new documents, or revisions of
their existing documents. 
A distinguishing feature is that the site strongly encourages users to
provide the source files for a document, rather than the ``compiled''
version. 
In particular, if a PDF generated from \TeX/\LaTeX\ is detected, it
is rejected, and the user is requested to provide the source files
instead. 

Several submission formats are allowed, including 
\TeX/\LaTeX, HTML, PDF, Postscript and (MS) Word. 
Our study focuses on the computer science and mathematical domains,
and (as we see below) in these areas, \TeX/\LaTeX\ predominates, and so
forms the bulk of our discussion\footnote{For convenience, in what
follows we refer to \latex, with the understanding that this
incorporates the \TeX\ format.}.

\paragraph{Data collection from \arxiv.}
In addition to a conventional web interface,
\arxiv\ provides an API for access to the data\footnote{\url{http://arxiv.org/help/api/index}}, which
we used for our data collection. 
Papers are arranged into a curated hierarchy: for example, 
{\tt cs.AI} is the Artificial Intelligence category within Computer
Science. 
We collected all papers with the area of computer science, and a large
subset from the area of mathematics, as of April 2011. 
Some papers have multiple categories: a primary category, and possibly
some additional categories.  
Our data collection method captured each paper once under its primary
categorization. 

As of April 2011, the \arxiv\ listed a total of
39015 CS papers and 196573 Math papers under all categories, however
this double counts many papers with multiple labels. 
We collected a total of 65235 papers:
 26057 from CS, representing all unique papers, and 39178 from math.
For math, we picked an arbitrary subset of subcategories, and
collected all papers in these categories (specifically, this was the
set of subcategories ordered by their two character names in the range
{\tt math.AC} to {\tt math.MG}).

\paragraph{Data set.}
\arxiv\ presents six fundamental document formats: 
 the well-known portable document format (PDF) and postscript; 
 HTML and the open XML document format used by recent versions of
 Microsoft Office and other wordprocessors (docx); and two variants of
 \latex, x-eprint and x-eprint-tar. 
Here, x-eprint corresponds to a single \latex\ source file with no
other files (i.e. no additional files containing figures,
bibliographical data, other \latex\ input files); while x-eprint-tar
is a `tar' archive file that contains multiple files compiled with \latex.

\begin{table}\centering
\small
\caption{Dataset by filetype}
\subtable[File Types in Arxiv\label{tb:filetype}]{\rowcolors{1}{}{LightSkyBlue}
\begin{tabular}{l|c|c}
 	File Type& number of Papers &Ratio\\
	\hline
	pdf & 7860 & 12\%\\
	postscript & 526 & 0.8\%\\
	text-html & 124& 0.2\%\\
        docx  &151 &0.2\%\\
	x-eprint &	28533&44\%\\
	x-eprint-tar &	28042&43\%\\
\end{tabular}
}\qquad
\subtable[Filetypes by subject\label{tb:papers}]{\begin{tabular}{c|c|c}
 & CS & Math \\
\hline
x-eprint-tar& 14964 (82\%) & 13088 (34\%) \\
x-eprint & 3334	 & 25199  \\
Dates & 1/1993 -- 4/2011 & 1/1991 -- 4/2011
\end{tabular}
}
\end{table}

 Table~\ref{tb:filetype} shows the distribution of formats for our
 dataset. 
It is striking that within computer science and mathematics, 
the \LaTeX\ formats predominate: they cover over 87\% of all papers. 
Submissions in HTML and docx formats are negligible, totalling less
than 0.4\%.  
From the PDF files, we extracted the metadata fields of ``title'',
``producer'' and ``creator''. 
Studying these indicates that a majority of this PDF files were in
fact created with the Microsoft Word software: 70\% of PDFs contain ``Microsoft''
or ``word'' in these fields. 
We note that docx is a relatively new format, and that as of July 2011
the \arxiv\ no longer accepts docx submissions, due to
difficulties with font conversions. 
Instead, papers written non-\latex\ tools are encouraged to be
submitted in PDF format.  

The \arxiv\ contains papers in Math and CS going back almost two
decades: papers in Math are indexed back to 1991, and in CS to 1993.
Table \ref{tb:papers} shows the breakdown of \LaTeX\ types by the two
major subject areas studied.
There is already a striking disparity between the two styles: a majority of
Math submissions are contained within a single \LaTeX\ file, while a
large majority of CS papers are spread across multiple files.

\begin{figure*}
\centering
\subfigure[Fraction of papers added each year\label{fig:paperyear}]{
\includegraphics[width=0.33\textwidth]{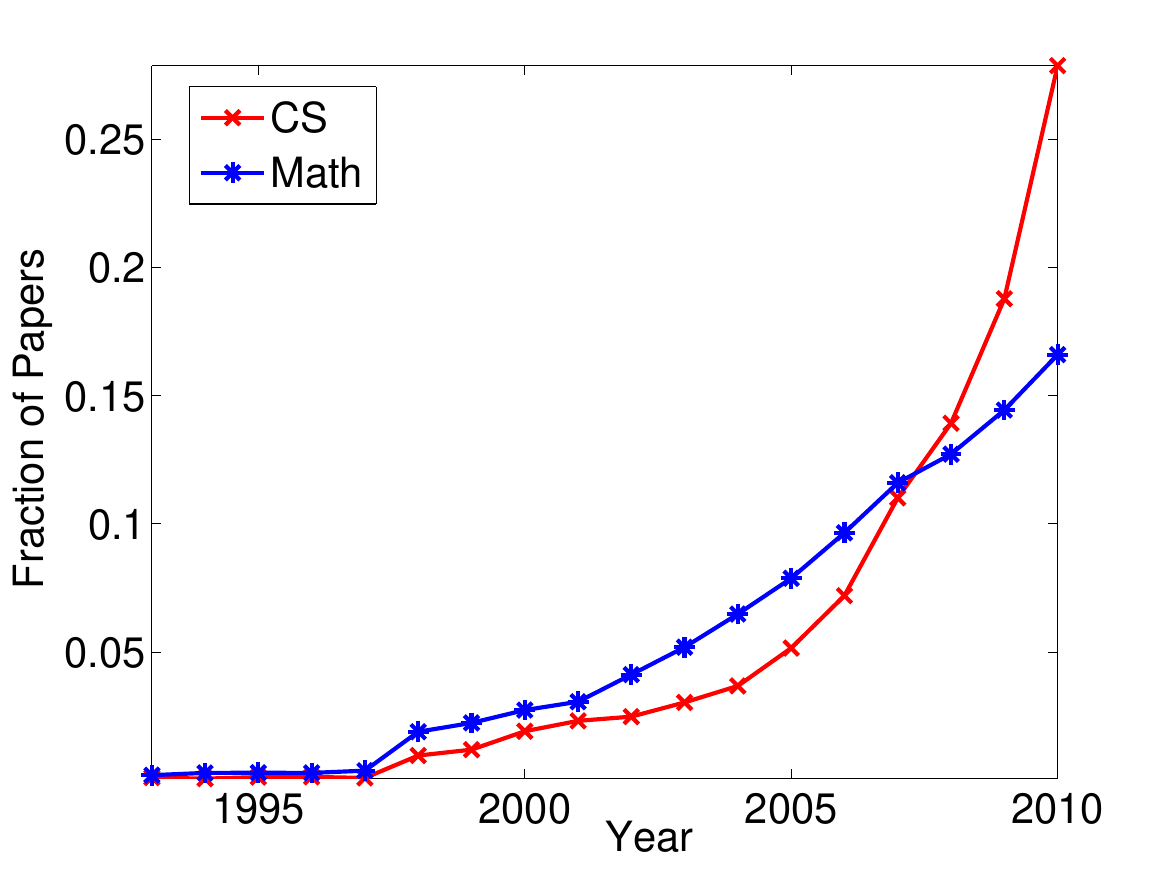}
}\subfigure[CS Paper Submission in each Month]{
		\includegraphics[width=0.33\textwidth]{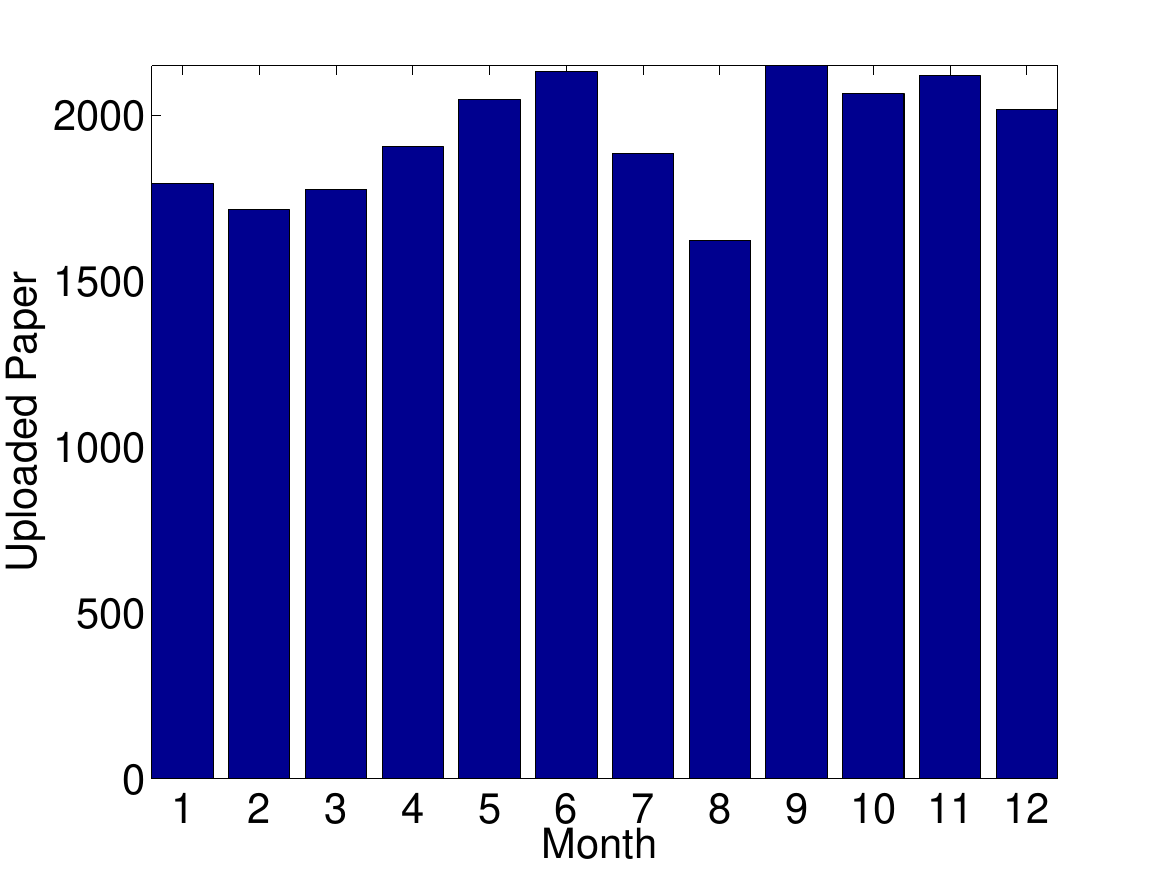}
		\label{fig:bymonthcs}
		}\subfigure[Math Paper Submission in each Month]{
		\includegraphics[width=0.33\textwidth]{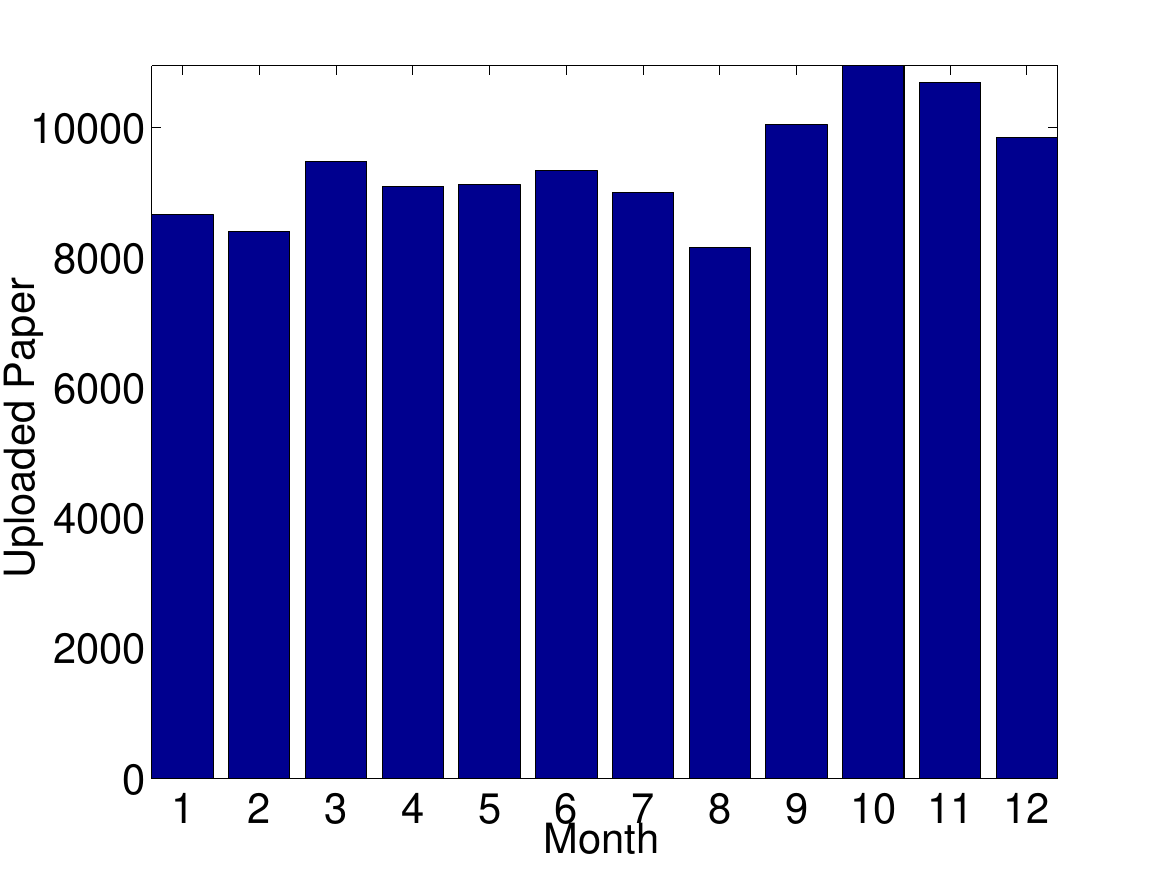}
		\label{fig:bymonthmath}
		}
  \caption{Paper submissions over time}
	\label{fig:bymonth}
\end{figure*}

Each paper is timestamped with the date of its upload.  
arXiv already shows some basic statistics on month-by-month
submissions for each field in its web interface\footnote{i.e.  for
  mathematics, see \url{http://arxiv.org/archive/math}}.
Figure~\ref{fig:paperyear} shows the fraction of papers
in each year for computer science and mathematics.
The trend for both areas is clearly increasing over time, with an
accelerating trend for CS while the growth in Math appears to be
increasing linearly year-on-year. 
We plot the histogram for uploaded  papers in each
month for both subjects in Figures \ref{fig:bymonthcs} and
\ref{fig:bymonthmath}. 
There is a clear lull in submissions around August and July, which
corresponds to the ``summer break'' in many (northern hemisphere) 
academic institutions.
We leave it to readers to conjecture explanations for this variation. 
Anecdotally, it is said that the summer months are used by researchers
to perform new research. 
This may be consistent with the figures if we accept that the fruits
of this research may not result in papers ready for submission until
some months later.  
Certainly, 
for mathematics, October and
November are months when people are most likely to submit papers to
arXiv, while June and September have the highest volume of
submissions in CS. 

\section{Structure analysis}
\label{sec:analysis}
Having access to scientific papers in \latex\ format enables us to
perform analysis which is either impossible or very challenging when
working with ``output'' formats such as PDF. 
For example, \latex\ files contain comments which are not present in
the final output, and identify the packages (libraries) used, which is
hard to do just by examining the output. 
We also want to study the use of expository structures 
like figures and theorems in scientific writing. 
While it is possible to identify these within PDF output, building
tools to do so is difficult, due to the many ways they can appear, and
the need to avoid false positives. 
In the \latex\ source, it is typically easier to identify these
structures, since the input is plain text, and there are only a few ways to
include such structures.

\subsection{Comments}
\label{sec:comments}
Much like a programming language, \latex\ allows {\em comments} within
the source files, which are ignored by the compiler and so
 do not appear in the final output of the paper (such as PS or PDF).
As such, they have the potential to shed extra light on the process of
writing science by capturing internal communications between authors, 
notes, earlier drafts or excised sections.  
Based on our inspection of the data set, we identified the following
usages of comments:

\paragraph{Templates and outlines.}
A basic use of comments is to provide an outline of the structure of
the paper, either as a reminder for the authors, or to help in the use
of a \latex\ template from a publisher. For example, serving as outline: 
{\scriptsize
\begin{verbatim}
	%% --------------------------Beginning of journal.tex----------------------
\end{verbatim}
}
describing the usage of latex command or package :
{\scriptsize
\begin{verbatim}
	% Use \tbl{...} command for table caption i.e. to fit table width.
\end{verbatim}
}
\paragraph{Internal communication.}
Some comments are used for communication between authors, such as:
{\scriptsize
\begin{verbatim}
%[xixi: Does it make sense now, as I can't find any direct reference]
\comment\{Should we have a short introduction paragraph to say what we're about to talk about? - Andy}
%Reviewer: Is there a range for $\phi$? I'm not sure what you mean by in that last sentence
\end{verbatim}
}
Some authors write hints or notes in comments to remind
himself/herself. For example,  
{\scriptsize
\begin{verbatim}
%% Requires GNUPLOT, compile with  ``pdflatex --shell-escape'' for the plots.
%TODO: needs more explanation!...
\end{verbatim}
}
\paragraph{Removed text.}
Many comments are just abandoned words, sentences or paragraphs which
are removed by authors.
For example, 
{\scriptsize
\begin{verbatim}
	%% \item If two walks that end at the same vertex induce different paths, output REJECT. 
	%% \item Consider two walks that end at adjacent vertices $u,v \in S$. If the walks together
	%%  with the edge $(u,v)$ give a cycle, output REJECT.  
	%% \item If no such cycles are detected, output ACCEPT.
\end{verbatim}
}

\eat{
\paragraph{Internal communication.}
Some comments are used for communication between authors, such as:
{\scriptsize
\begin{verbatim}
%added by Jarek
%[xixi: Does it make sense now, as I can't find any direct reference]
%[xixi:Is it ok to say this?]organization
%[TODO: justify the ... model as an appropriate basis for
 a simulation of semantic-social networks, xixi: is it enough?]
\comment\{Should we have a short introduction paragraph to say what we're about to talk about? - Andy 
Reviewer: Is there a range for $\phi$? 
Tree Animation Example Animation Example p is not defined!!
I'm not sure what you mean by in that last sentence
\end{verbatim}
}
Some authors write hints or notes in comments to remind
himself/herself. For example,  
{\scriptsize
\begin{verbatim}
%% Requires GNUPLOT, compile with  ``pdflatex --shell-escape'' for the plots.
%TODO: needs more explanation!...
\end{verbatim}
}
\paragraph{Removed text.}
Many comments are just abandoned words, sentences or paragraphs which
are removed by authors.
For example, 

{\scriptsize
\begin{verbatim}
	%% \item If two walks that end at the same vertex induce different paths, output REJECT. 
	%% \item Consider two walks that end at adjacent vertices $u,v \in S$. If the walks together
	%%  with the edge $(u,v)$ give a cycle, output REJECT.  
	%% \item If no such cycles are detected, output ACCEPT.
\end{verbatim}
}
}
We begin this study by studying the prevalence and basic
characteristics of comments. 
In \latex, there are a variety of methods to add comments in an article. 
The principle methods are:

\begin{enumerate}
\item the built-in latex comment command: `\%'
\item use \textsf{ \textbackslash newcommand} to define a function
  that ignores its parameter, as
  \textsf{\textbackslash newcommand\{\textbackslash comment\_tag\}[1]\{\}}  
\item define conditional comments such as 
\textsf{\textbackslash newcommand\{\textbackslash condcomment\}[2]\{\textbackslash ifthenelse\{\#1\}\{\#2\}\{\}\}}
\item use commands in special packages, such as packages {\em verbatim}
  and {\em comment}
\end{enumerate}

We manually checked a large sample of papers and found the first two
were by far the most common methods used. 
Therefore, we built scripts using regular expressions to detect their usage.
For the first case, any string after the special character `\%' 
and before the next newline are extracted as comments.  
Similarly, for the second case, we identify all instances of commands
``comment\_tag'' that ignore their parameter, and then extract all
subsequent text within a ``comment\_tag'' environment.

The advice on \arxiv\ to authors uploading their papers is to remove
comments from their submissions\footnote{See \url{http://arxiv.org/help/faq/whytex}}.  
However, the above procedure found comments in 90.4\% of Math papers
and 95.3\% of CS papers. 
In many cases, the comments remaining are minimal or innocuous;
however we also saw many examples of the form described above, which
might be considered sensitive by the authors. 
For CS papers, the average number of words in comments per paper is 772; 
for math, it was 395. 
Expressed as a percentage of the total length of papers, this
corresponds to 
 7.2\% in CS, and 3.9\% for in math. 
There is an appreciable difference in vocabulary size: 
 in the full papers, there are around 1.3 million distinct words in
 CS, and 1.5 million in Math papers. 
Restricting attention to just the comments though, 
there are only 299 thousand distinct words in
CS papers, and 338 thousand for math. In percentage, comments consume 15\% words for CS and 18.7\% for math.

\begin{figure*}[t]
\vspace*{-0.5in}
\subfigure[Comments in 2008]{
  \includegraphics[width=3.5in,angle=0]{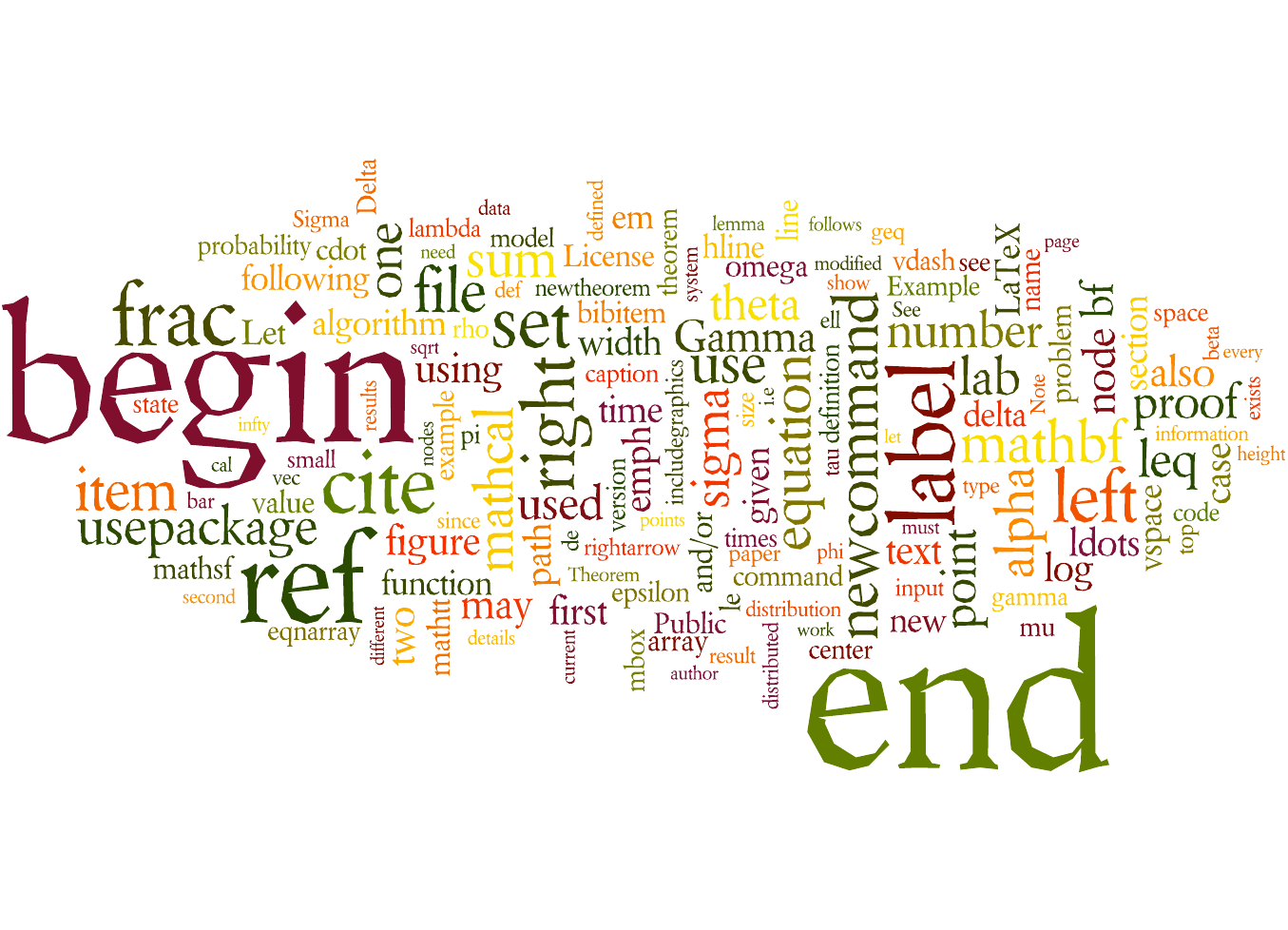}
  \label{fig:wordle08}
}\subfigure[Comments in 2009]{
  \includegraphics[width=3.5in,angle=0]{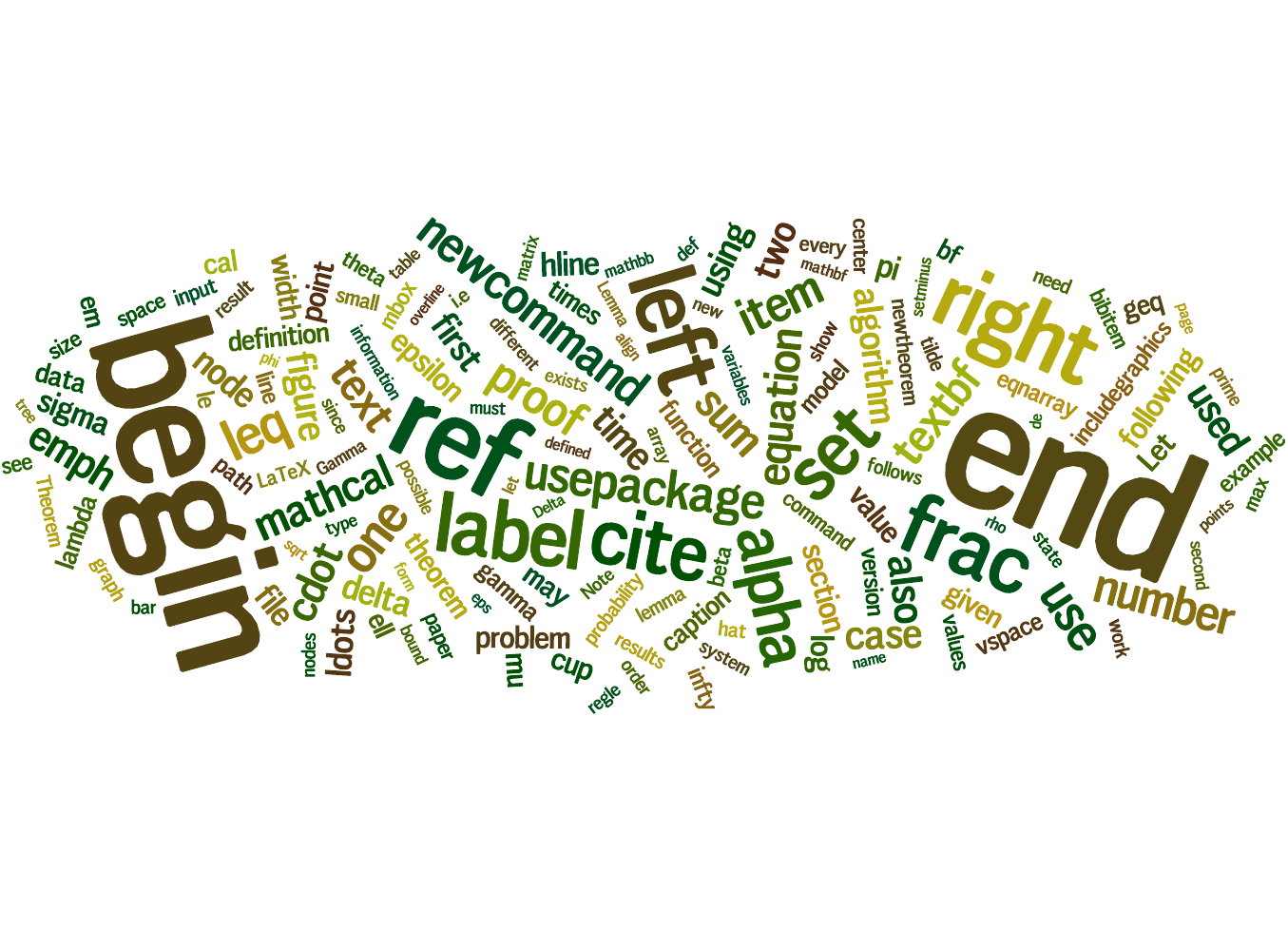}
  \label{fig:wordle09}
}
\caption{Wordcloud of Comments in papers of 2008 and 2009}
\end{figure*}

\begin{figure*}[t]
\vspace*{-0.5in}
\subfigure[Comments in 2008]{
  \includegraphics[width=3.5in,angle=0]{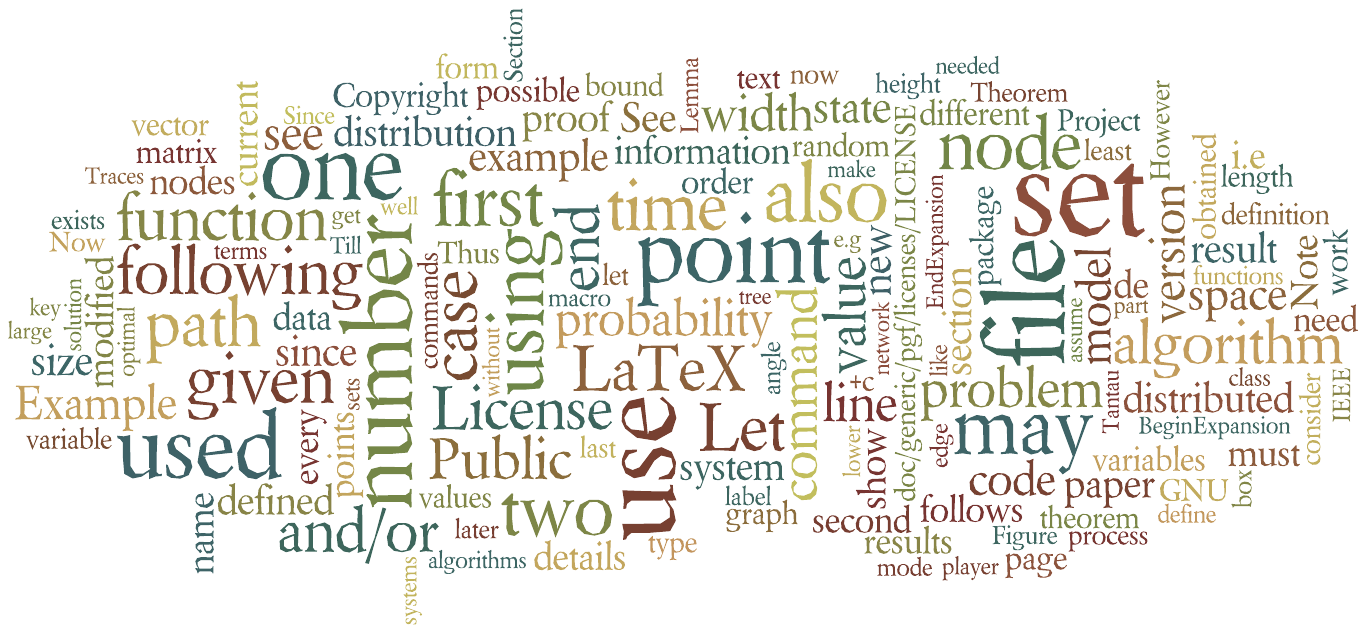}
  \label{fig:wordle08nolatextag}
}\subfigure[Comments in 2009]{
  \includegraphics[width=3.5in,angle=0]{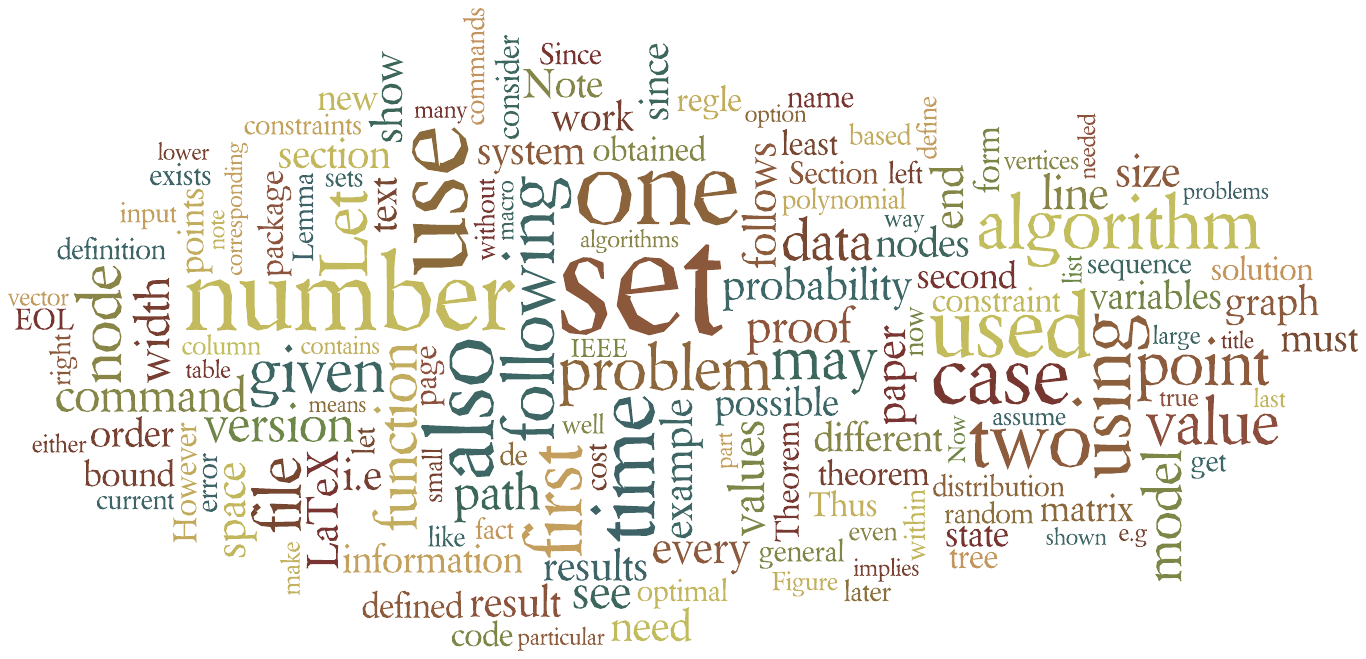}
  \label{fig:wordle09nolatextag}
}
\caption{Wordcloud of Comments without latex command tags in papers of 2008 and 2009}
\end{figure*}
To generate a visualization of the vocabularly of comments, we used
the Wordle tool\footnote{\url{http://wordle.net}} to
generate a word cloud of comments. 
Figure \ref{fig:wordle08} shows the frequent words in comments from CS
papers from 2008, while Figure \ref{fig:wordle09} is the corresponding
plot for 2009. %Removing  \latex-specific commands such as '$\backslash$begin\{equation\}', the corresponding plot for the year 2008 is in Figure \ref{fig:wordle08nolatextag}  and for the year 2009 is in Figure \ref{fig:wordle09nolatextag}.
Here, the size of each word scales with to the frequency of the word
in the input. 
For clarity, short words (of 1 or 2 characters) are removed, as are
standard English stop-words (using the list in appendix 11 of
  \cite{stoplist}).

By visual inspection, it is clear that \latex-specific commands
are very significant in comments: ``begin'' and ``end'' control the
scope of environments, while ``ref'', ``cite'' and ``label'' manage
references within papers. We manually identify two dominant clusters of terms. 
One cluster has \latex\ formatting terms, such as  
\begin{center}
{\em figure, ldots, mbox, end, begin, label, 
cite, newcommand, item, section}
\end{center}
The other cluster has terms related to mathematical expressions, such as
\begin{center}
{\em frac, equation, left, right, mathcal, leq, alpha, 
delta, sigma, phi, gamma, beta, omega, sum}
\end{center}

\eat{We manually identify two broad clusters of terms. 
One cluster has terms related to mathematical expressions,
such as  
\begin{center}
{\em frac, equation, left, right, mathcal, leq, alpha, 
delta, sigma, phi, gamma, beta, omega, sum}
\end{center}
The other cluster has \latex\ formatting terms, such as 
\begin{center}
{\em figure, ldots, mbox, end, begin, label, 
cite, newcommand, item, section}
\end{center}
}
We can consider various hypotheses for this behavior. 
It may be that these words are globally frequent across
\latex\ papers, both within and outside of comments. 
Or it may be that these are more popular in comments, since 
writing Math and \latex\ formatting is considered more tricky, and so
authors retain different commands in comments, whereas redrafts of
text are just overwritten.  

When we compare the word frequency distribution between comments and
the rest of the papers, we do not observe a very large difference. 
However, there are some words which are more common in comments than
in the rest of papers, and vice-versa.  
We can find those words which have the largest absolute change in
(normalized) frequency between two inputs.  
The 10 most discriminative words of comments in CS papers compared to
the remainder of those papers, in descending order, are

\smallskip
\centerline{\em latex, tex, file, use, usepackage, you, end, sty, text,
  version.}

\smallskip
\noindent
whereas, in the opposite direction, the top 10 discriminative words of
CS papers comparing to their comments are 

\smallskip
\centerline{\em equation, let, each, one, def, sec, two, model, function, given.}

\smallskip
The presence of ``you'' (and, more ambiguously, ``use'' and
``version'')
in comments strongly suggests the importance
of comments for communication between authors.  
In contrast, the words that are discriminative for the text seem to
mostly relate to more formal computer science writing. 

The top 10 discriminative words of comments in Math papers comparing
to the rest of the papers are

\smallskip
\centerline{\em
tex, latex, file, end, math, macros, text, use, version, line}

\smallskip
\noindent
while the top-10 discriminative words of Math papers compared to their
comments are

\smallskip
\centerline{\em 
let, equation, such, where, theorem, proof, have, lemma, follows, proposition}

\noindent
which again appears to show a difference in the use of comments than
for the main text. 

\medskip
\noindent
{\bf Defining and finding comments.} 
We have been somewhat quick in defining the concept of ``comments'' thus
far, in the interest of adopting a workable definition for our empirical study. 
For a more formal notion, 
denote the input string  as $s=s[1,n]$, where each $s[i] \in \Sigma$
for some set $\Sigma$ of symbols, and assume 
a function (program) $P: \Sigma^* \rightarrow \Sigma^*$ that maps input
strings to output strings. 
We can now give a semantic definition: a {\em comment} is a
substring $s[i,j], i \leq j$, such that $P(s) = P(s[1,i-1]s[j+1,n])$.
In many applications, we can assert that if
$s[i,j]$ is a comment, so is a substring $s[k,l], i \leq k \leq l
\leq j$.  
To make such comments a  
semantic unit, we define a {\em maximal comment} as a substring
$s[i,j], i \leq j$, such that 
\[ P(s) = P(s[1,i-1]s[j+1,n]); \quad P(s) \not=P(s[1,i-2]s[j+1,n]) 
; \quad 
\text{ and } P(s)\not=P(s[1,i-1]s[j+2,n]).\] 

Note that maximal comments do not overlap. Using an oracle that will
check if $P(t_1) = P(t_2)$ for two strings $t_1, t_2$,
we can now address questions of interest such as,  
(a) Is $s[i,j]$ a comment? (b) Is $s[i,j]$  a maximal comment, and (c)
What is a partition of $s$ into maximal comments?, and find
efficient algorithms.  

Mapping this problem to the \latex\ case provides further questions. 
In the simplest mapping, $s$ is a \latex\ document viewed as a
sequence of symbols, $P$ is the \latex\ compiler, and the output
is the pdf version (say).  
We assume that the output does not change when a comment is removed,
and substrings $s[i,j]$ whose removal makes  $s[1,i-1]s[j+1...n]$
illegal for the compiler can be detected.
However, this definition means extra whitespace is treated as comments. 
A  more \latex-aware way to do the mapping is to consider only parsed
``words'' that arise from \latex\ language, and treat them as symbols.  
Then the \latex\ document is viewed  
as the rooted {\em hierarchy} of environments which can be thought of
as a tree. 
Here, the formal concepts still apply at every level of
such a tree, treating symbols and nodes suitably.
Finally, we can imagine   
simulating the
\latex\ compiler, keeping its state, and detecting comments online
during processing.   
A last open question is, can one formally model 
and prove that a  general  \latex\  compiler needs more
resources --- space, time, passes --- than that needed to detect
comments.  
\qed

\subsection{Length}
A fundamental property of research papers is their length: how long
does it take a researcher to articulate their novel ideas?  
How does this vary across areas, and across time?
In \arxiv, the length of the paper is metadata provided by the
uploader. 
Based on a spot check, we established that this is typically accurate.
However, in some cases, this optional information is omitted. 
In other cases, where multiple versions of a paper are uploaded, only
the length of an initial version is provided, and is not corrected as
new versions are provided. 
Within our dataset, 26\% of CS papers and 45\% of Math papers have
page number information.

\begin{figure*}
\centering
\subfigure[Page number distributions for Math and CS]{
\includegraphics[height=0.27\textwidth,width=0.33\textwidth]{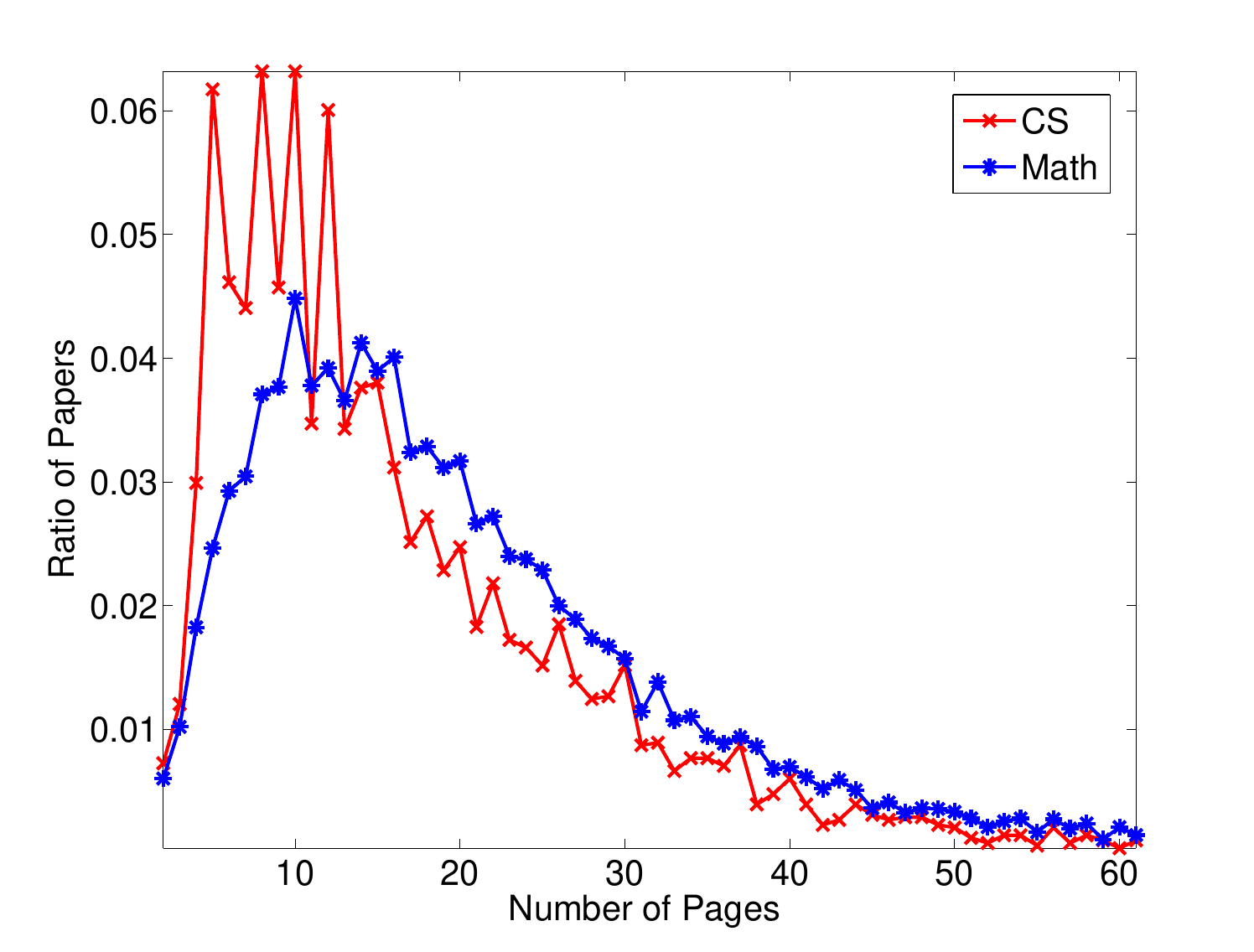}
\label{fig:pagenumbers}
}\subfigure[Pagelength over time]{
\includegraphics[height=0.27\textwidth,width=0.33\textwidth]{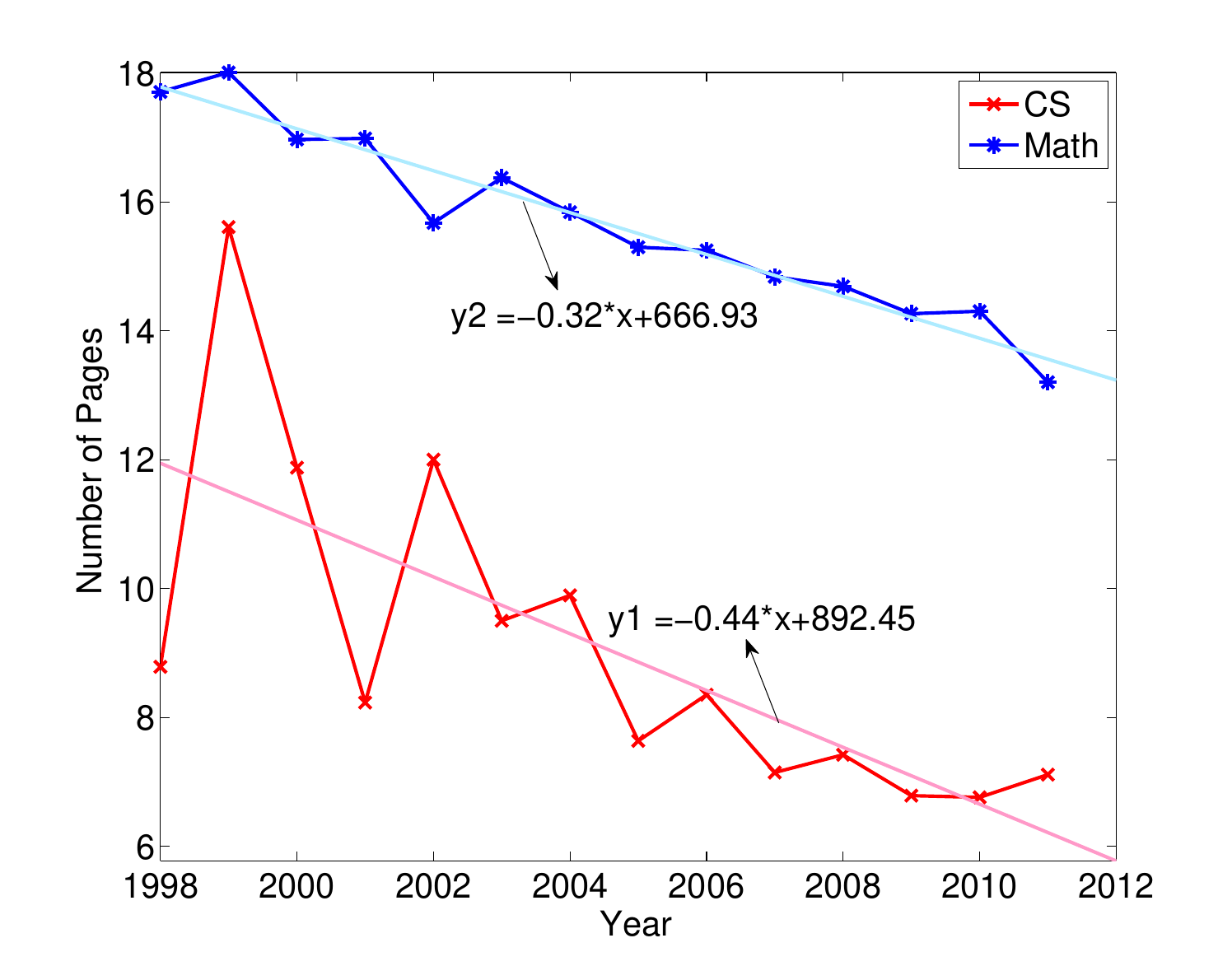}
\label{fig:page}
}\subfigure[Wordlength over time]{
  \includegraphics[height=0.27\textwidth,width=0.33\textwidth]{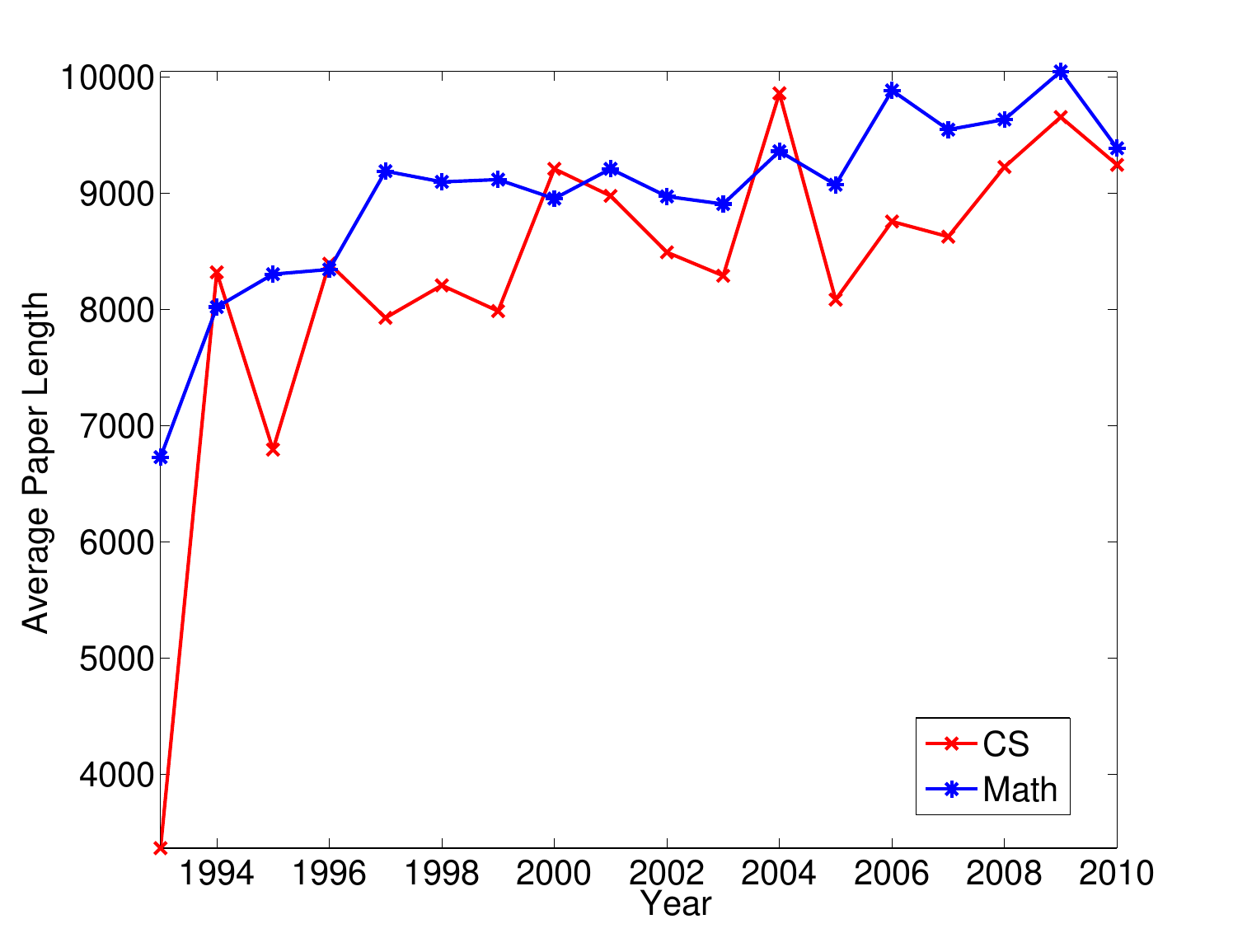}
  \label{fig:wordtime}
}\caption{Length trends over time}
\end{figure*}

Figure~\ref{fig:pagenumbers} shows the page number distribution of both
subjects. 
The difference between the two distributions is quite striking. 
Math follows an approximately unimodal distribution with a peak around 10
pages.  
For CS, there are multiple peaks which seem to alternate page
lengths. 
Our hypothesis is that this corresponds to submissions to
conferences that had been uploaded to the \arxiv: conference page
limits are typically around ten pages.  
Indeed, the observed peaks occur at 5, 8, 10 and 12 pages, all of
which are page limits for various conferences.  
There is a slight preference for papers with an even number of pages,
but not excessively so: 52\% of Math papers have even length, and 54\%of CS papers. 
The average length of Math papers is slightly greater than that of CS
papers: 
9345 words compared to 9011 words. 
However, the difference in page lengths is more appreciable, averaging
15 pages in Mathematics to 9 pages in CS. 
This suggests a tendency to use denser page layouts in CS.

As time goes on, do papers get longer, as more concepts and related
work need to be explained?
The trend actually seems to be the reverse, as shown in 
Figure~\ref{fig:page}. 
The behavior of Math in particular seems to be well-fitted by a linear
trend, removing 1/3 page per year. 
Extrapolating this line beyond the bounds of common sense, we conclude
that the average Math paper will have no pages by the year 2052. 
For computer science, this date will be 2026.   
However, when we view the length of papers in terms of words (Figure~\ref{fig:wordtime}), we see
that the trend is upwards. 
So we conjecture that the real behavior is that more papers are being
posted to \arxiv\ in dense layouts, which pack more words per page.

\subsection{Package Use}
\label{sec:packages}
\begin{table}
\centering
\caption{Top 20 Used Packages\label{tb:top30pack}}
\small
{\rowcolors{1}{}{LightSkyBlue}
\begin{tabular}{c|l|l||l|l}
Rank & CS packages & Fraction &Math packages&Fraction\\
\hline
1 & amsMath & 0.52 & amssymb & 0.57\\
2 & amssymb & 0.51 & amsMath & 0.45\\
3 & graphicx & 0.50 & amsfonts & 0.27\\
4 & amsfonts & 0.22 & amsthm & 0.24\\
5 & epsfig & 0.21 & graphicx & 0.21\\
6 & latexsym & 0.18 & amscd & 0.17\\
7 & url & 0.18 & latexsym & 0.17\\
8 & color & 0.17 & xy & 0.14\\
9 & amsthm & 0.15 & epsfig & 0.10\\
10 & subfigure & 0.13 & color & 0.07\\
11 & times & 0.12 & mathrsfs & 0.06\\
12 & inputenc & 0.09 & inputenc & 0.06\\
13 & cite & 0.08 & hyperref & 0.05\\
14 & algorithm & 0.08 & enumerate & 0.05\\
15 & algorithmic & 0.08 & babel & 0.05\\
16 & hyperref & 0.08 & graphics & 0.05\\
17 & graphics & 0.07 & verbatim & 0.04\\
18 & fullpage & 0.06 & fontenc & 0.04\\
19 & xspace & 0.06 & eucal & 0.04\\
20 & babel & 0.06 & url & 0.03
\end{tabular}

}
\end{table}

\latex\ is a very flexible and extensible typesetting system. 
Additional functionalities can be added by making use of ``packages'',
via the \textsf{\textbackslash usepackage} command.
Each package implicitly captures the fact that authors need a
certain kind of styling or presentation to
better express their research. 
Hundreds of packages are available.
We extracted all the packages invoked in our dataset:
1480 distinct packages in CS and 988 in Math.
Table~\ref{tb:top30pack} shows the names of the top 20 most frequent
packages in each subject, and the ratio of papers that each appears in.

There are many commonalities to the list.  
Extensions provided by the American Mathematical Society (AMS) are the
most popular, providing commonly used mathematical symbols and
structures like theorems (amsmath, amsymb, amsthm). 
Other packages deal with including figures (graphics, graphicx)
and changing font family and color (times, color, fontenc). 
These functionalities are commonly needed in both Math and CS. 
However, there are notable differences between the usage patterns of
the two areas.  
For example, `algorithm' and
`algorithmic' included by many CS papers don't appear in top 20
packages of Math.
In Math papers, additional AMS packages are used that are not common
in CS. 

We found the most discriminative packages between the two areas (the
packages that have the largest absolute difference in usage). 
The top-10 discriminative packages for CS are

\centerline{\em
graphicx, url, epsfig, subfigure, times, 
color, algorithm, algorithmic, amsmath, cite}

The top 10 discriminative packages for Math are 

\centerline{\em amscd, xy, amsthm, amssymb, amsfonts, 
eucal, mathrsfs, xypic, amsxtra, euscript}

The packages discriminative for Math are all related to support for
certain symbols, fonts and diagrams common in Math that are rarely used in CS. 
For CS, the discriminative packages cover a broader range of uses:
referencing ({\em cite, url}), including and formatting figures ({\em graphicx,
epsfig, subfigure}), writing pseudocode ({\em algorithm, algorithmic}) and
styling text ({\em times, color}). 

\begin{figure*}[t]
	\centering
	\subfigure[Package use over time]{
		\includegraphics[height=0.3\textwidth,width=0.35\textwidth]{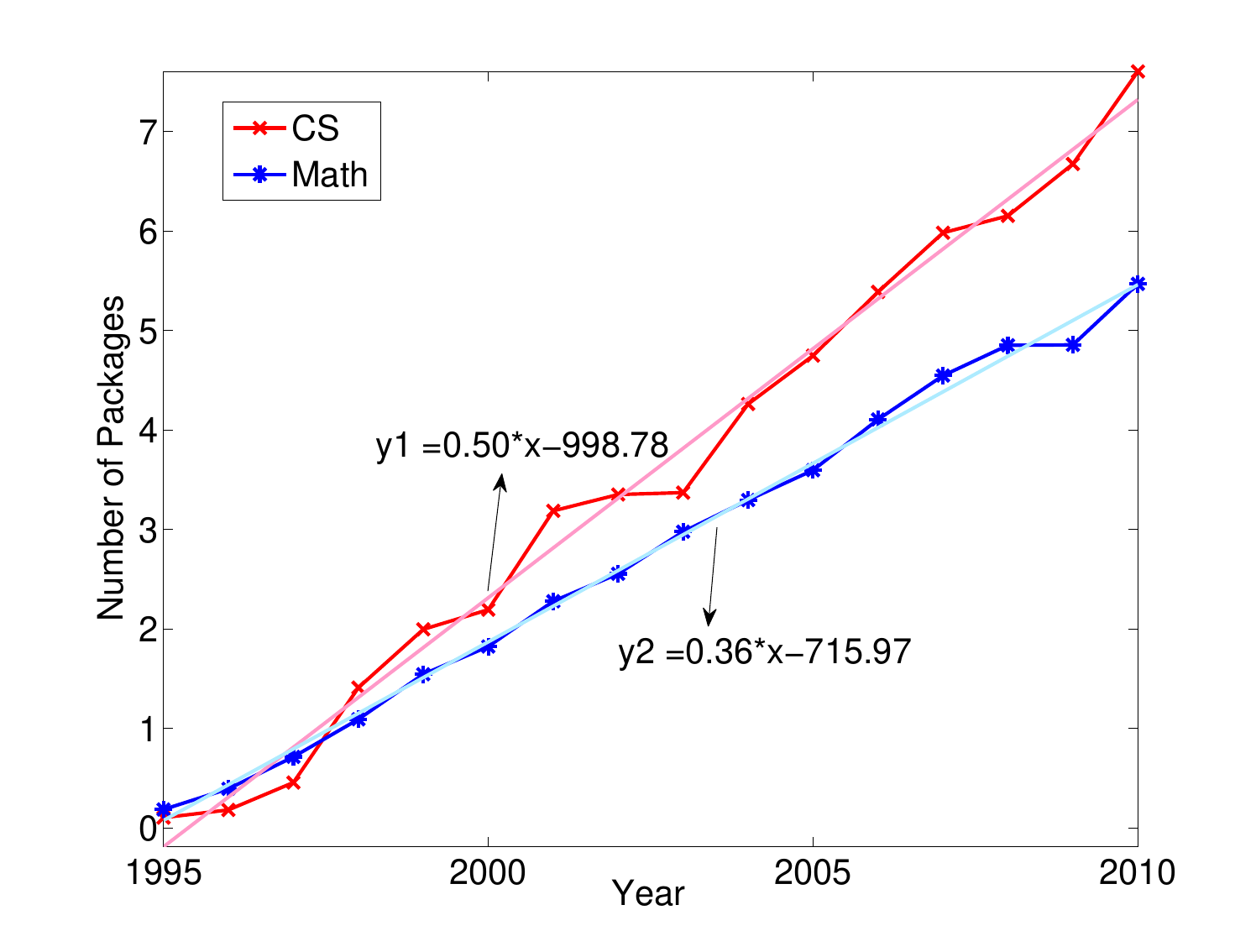}
		\label{fig:packbyyears}
		}	\subfigure[Paper length against number of packages]{
		\includegraphics[height=0.3\textwidth,width=0.35\textwidth]{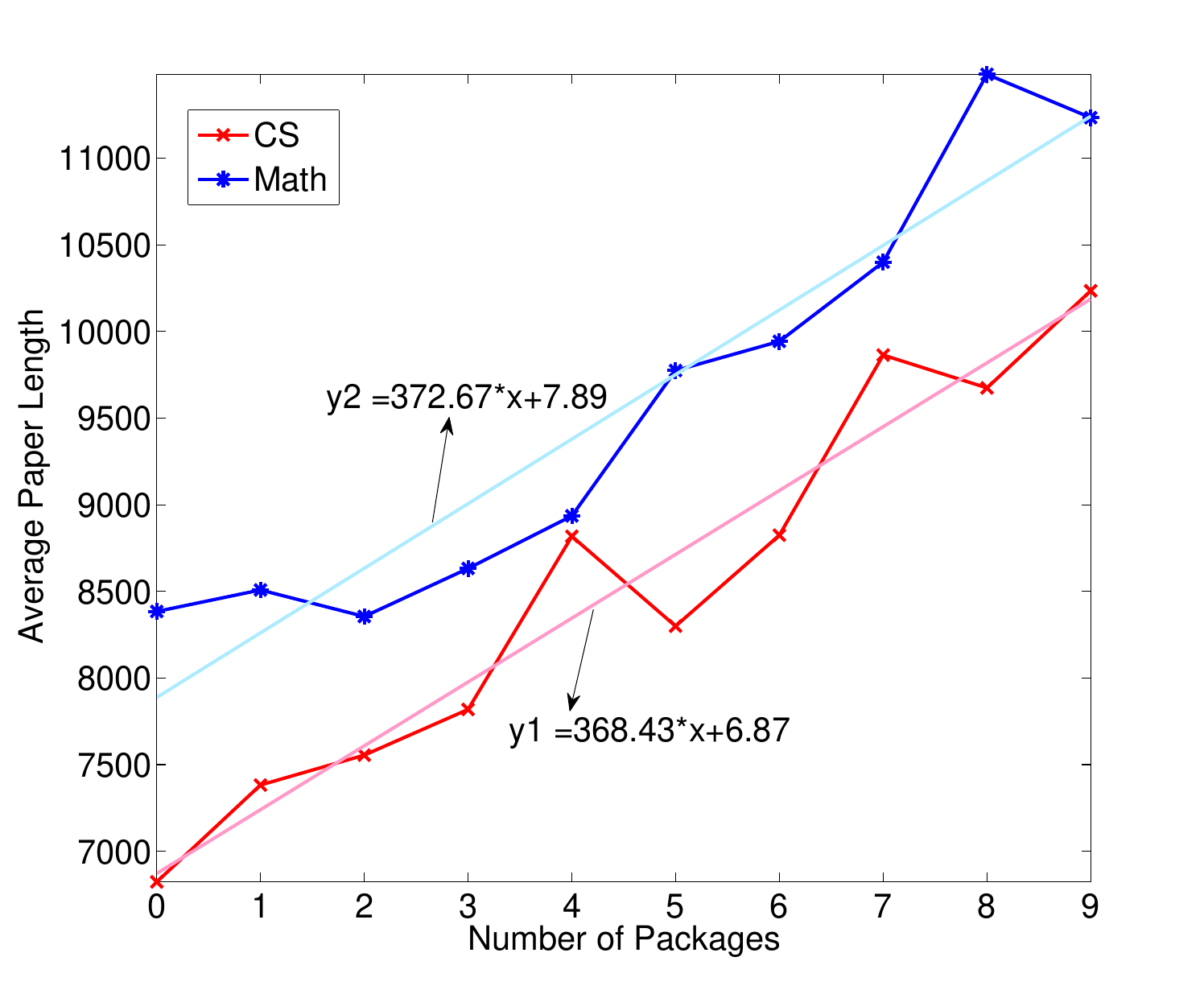}
		\label{fig:paperlen_pack}
		}
\caption{Package usage}
	\label{fig:package}
\end{figure*}

In CS, 87\% papers include at least one package; and the percentage is 75\% for math. %while for Math, 75\% papers have packages. 
Of those papers which do include packages, 
the average number of packages included in a paper is 6.7
for CS and 5.0 for Math. 
These numbers are close, but indicate a slightly greater need for
extra functionality in CS. 

Figure~\ref{fig:packbyyears} depicts how the average number of
packages per paper varies by year. 
In math, this growth is about 1 package every 3 years, while in CS it
is 1 package every 2 years.
This growth rate is indicative of changing needs of authors: 
\latex\ is relatively stable, and rarely adds features.
Yet increasingly authors need to access functionality provided by
packages, such as to include URLs and graphics files in their papers. 

When we plot paper length (in words) against the number of packages
used, we see a different effect in Figure~\ref{fig:paperlen_pack}. 
There seems to be an appreciable correlation between these two values,
and moreover this is very consistent between Math and CS: each package
seems to add 370 words to the paper. 
Perhaps a better way to view this is that as papers grow longer, 
they are more likely to require additional packages to help
express their ideas.

\begin{figure*}[t]
	\centering
	\subfigure[Two Packages: {\em graphicx} vs {\em epsfig}]{
		\includegraphics[height=0.3\textwidth, width=0.35\textwidth]{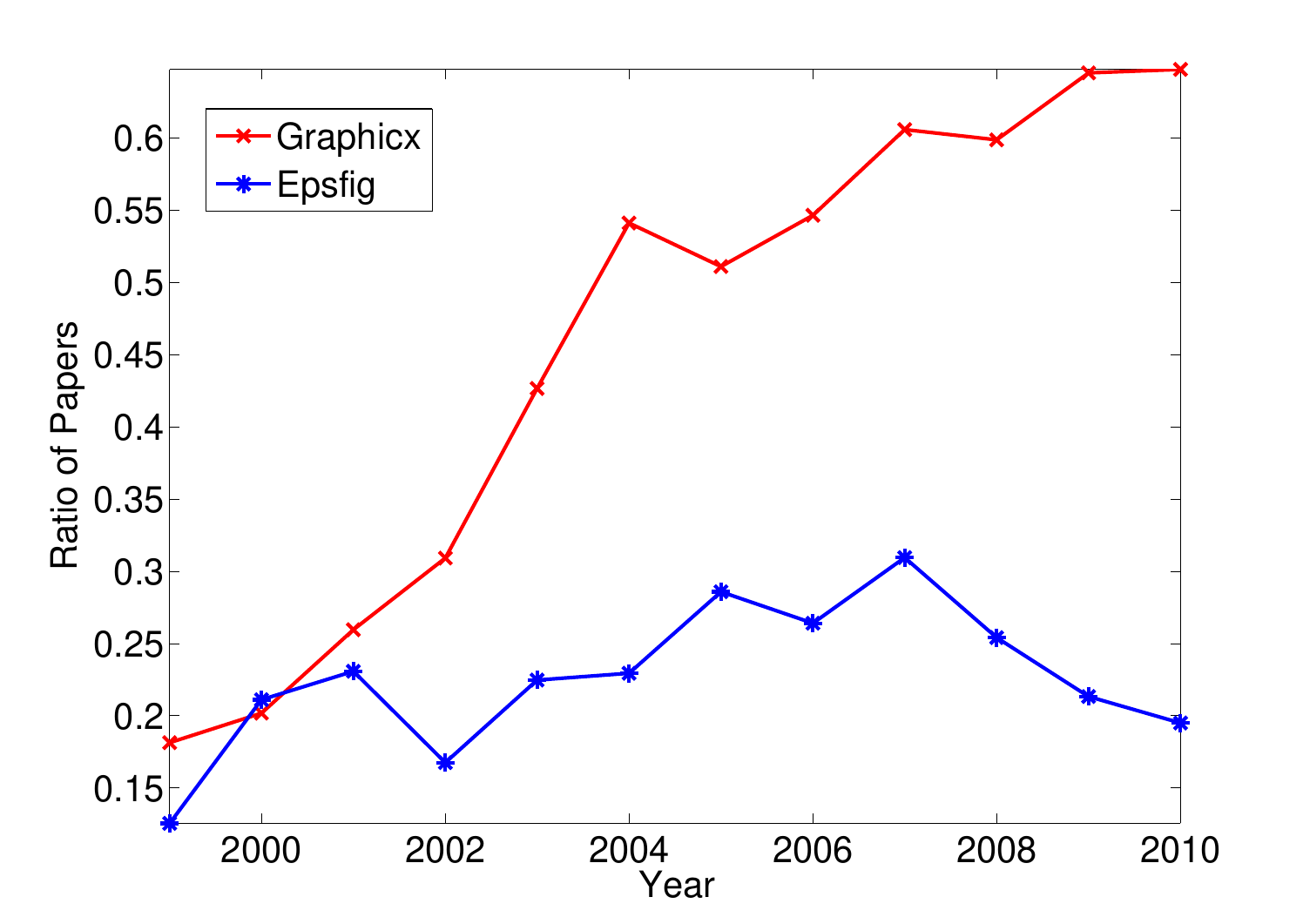}
		\label{fig:graphicxeps}
		}	\subfigure[Paper length against number of figures]{
		\includegraphics[height=0.3\textwidth,width=0.35\textwidth]{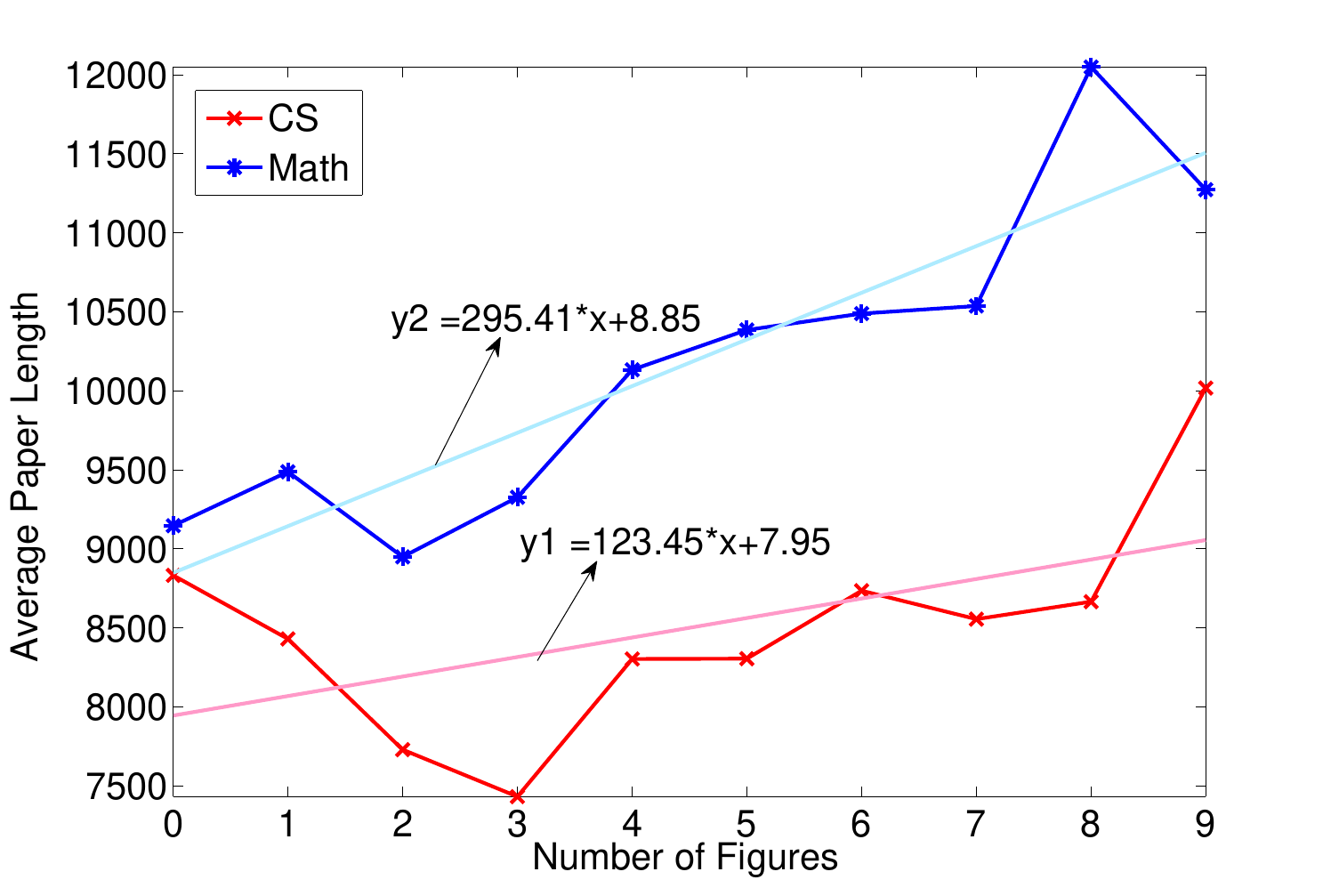}
		\label{fig:paperlen_fig}
		}
\caption{Figure usage}
\label{fig:fig}
\end{figure*}

\subsection{Figures} 
\paragraph{A tale of two packages: graphicx and epsfig.}
It is quite common to find multiple packages that provide very similar
functionality. 
One prevalent example is for including graphic files in
\latex\ documents. 
Epsfig\footnote{\url{http://www.ctan.org/pkg/epsfig}} 
allows the inclusion of figures in encapsulated
postscript (eps) format. 
It is generally considered deprecated, since it is not compatible with
modern (pdf)\latex\ compilers that output directly to PDF format. 
Instead,  graphicx is recommended, as it can handle a broader variety
of file formats\footnote{\url{http://ctan.org/pkg/graphicx}}.
However, both are widely used in our data set: graphicx is invoked in
more than half of CS papers, while epsfig appears in over 20\%. 
Figure~\ref{fig:graphicxeps} shows the usage trends of
both packages in CS. 
While it is clear that take-up of graphicx is increasing over time, 
it is not so clear that epsfig is receding: while the trend in recent
years seems downward, the trend over an 11 year period is upward. 

It is also of note that 10\% of papers invoke both packages, which is 
unlikely to be deliberate.  
On closer investigation, we observe some other odd behavior. 
For example, of the papers which call on graphicx, 28\% do not actually
include any graphics (i.e. they do not contain any
\textsf{\textbackslash includegraphics} commands). 
Even more pronounced, 84\% of paper which call epsfig do not in fact
use it (there are no  \textsf{\textbackslash epsfig} commands). 
We interpret this pecularity as informing us about the way in which
authors write their papers. 
The ``preamble'' of a \latex\ document, where packages are called and
macros are defined, can easily grow quite long.  
Faced with starting a new document, it may be easier to copy from an
existing paper and modify it rather than start afresh.
Some authors like to have the same macros defined in each paper they
write, and this is a quick way to achieve it. 
As a result, packages can be brought along which are not actually used\footnote{Rather
  fancifully, one can compare this process to evolution, and identify
  ``junk DNA'' with ``junk \latex''.}. 
This phenomenon may also partly explain the gradual growth in number
of packages. 
As a result, it seems that the epsfig package may continue to appear
in documents, even though it is not actually used within them. 

\paragraph{Testing folk wisdom.}
The old adage, ``A picture is worth a thousand words'', suggests that
adding illustrative figures should tend to reduce the length of a
document.  
However, figure~\ref{fig:paperlen_fig} shows the opposite trend: in
both Math and CS, adding figures {\em increases} the length of a paper. 
In math, the trend seems to be fairly consistent, and we have a new
adage: ``A pictures costs three hundred words''.
For CS, the trend is more variable, and weaker: the cost is an average of 120
words per figure. 
We might conjecture that in math, figures are typically illustrating
technical concepts which require some effort to describe, whereas in
CS, many figures are data plots that need less text to interpret.

\subsection{Number of Authors}

\begin{figure*}\centering
\subfigure[Words per author]{
  \includegraphics[height=0.3\textwidth,width=0.33\textwidth]{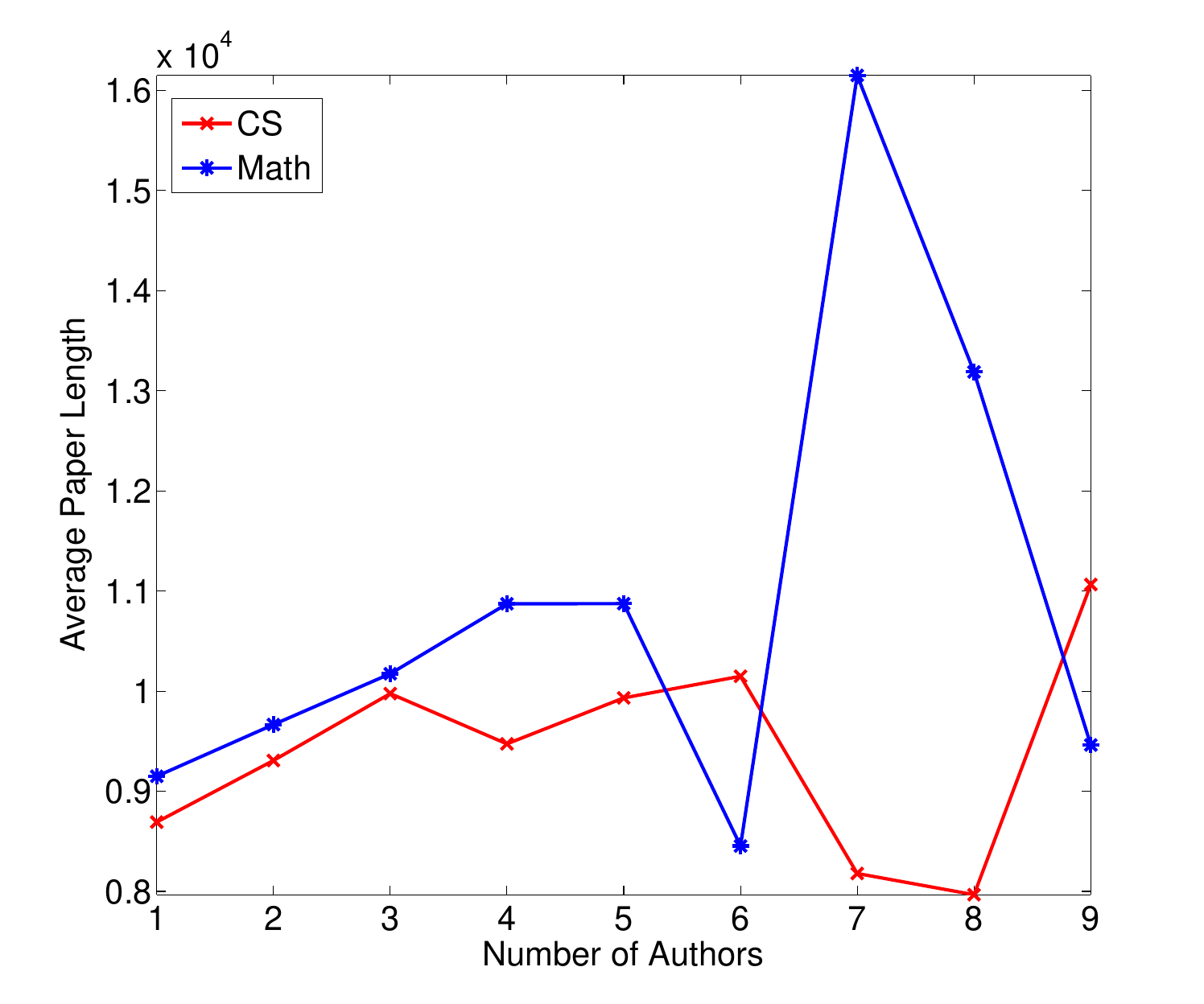}
  \label{fig:authortext}
}\subfigure[Comments per author]{
\includegraphics[height=0.3\textwidth,width=0.33\textwidth]{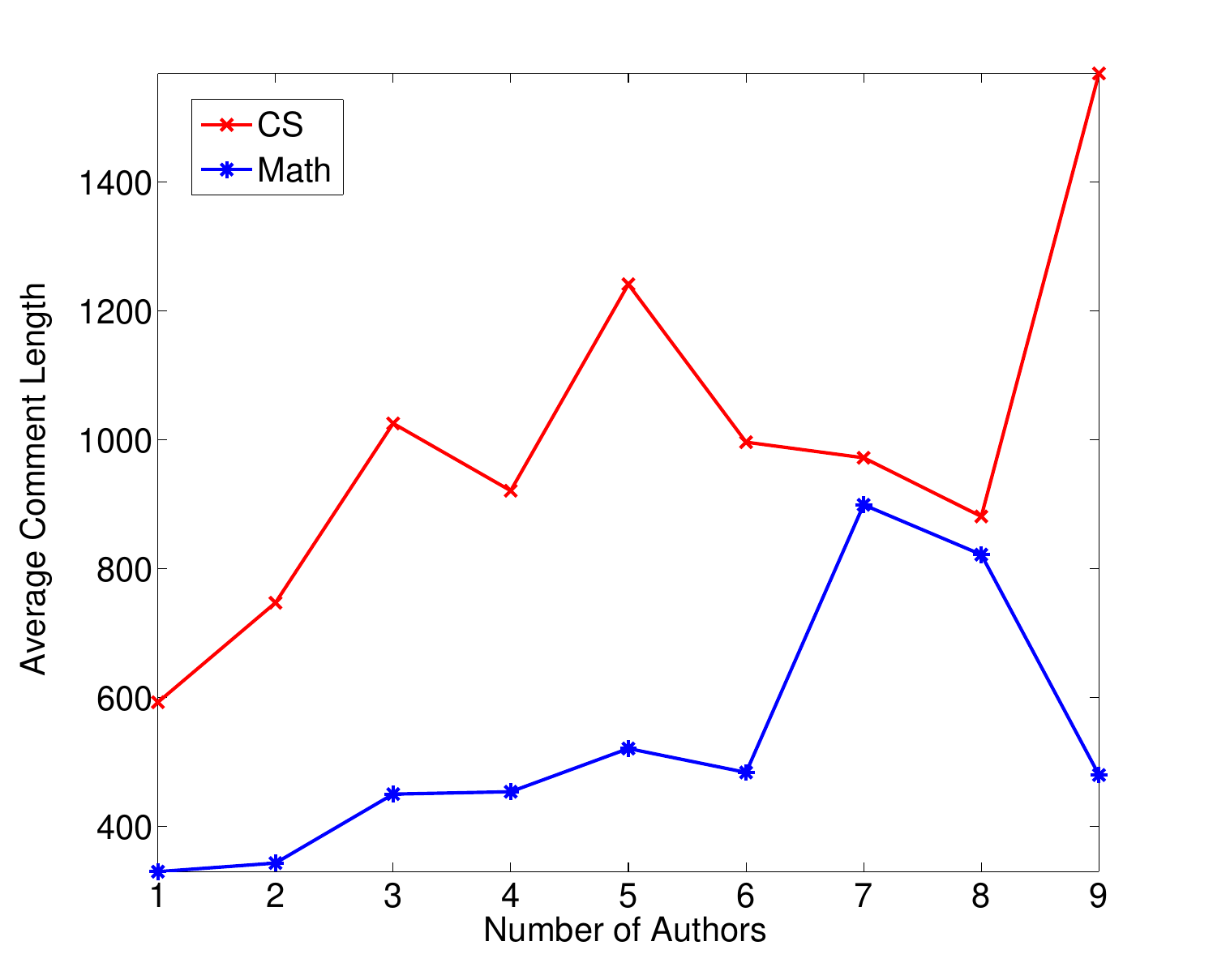}
\label{fig:authorcomments}
}\subfigure[Paper length against number of theorems]{
		\includegraphics[height=0.3\textwidth,width=0.33\textwidth]{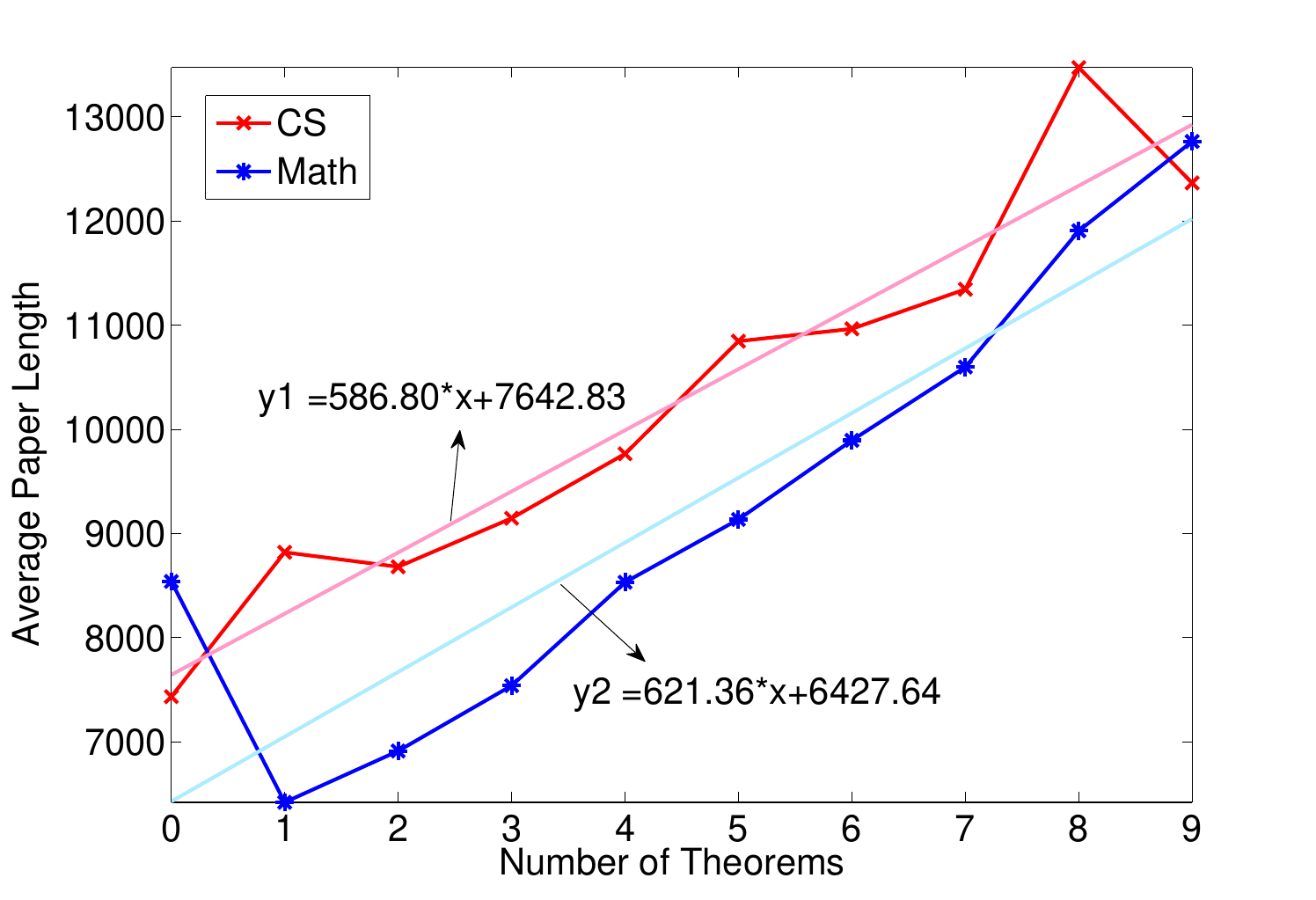}
		\label{fig:paperlen_thm}
		}

\caption{Length as a function of authors and theorems}
\label{fig:authorwords}
\end{figure*}

In some areas, it is common for multiple authors to jointly
collaborate on writing a paper. 
In this section, we study patterns of coauthorship within our data
set. 
The \arxiv\ API returns only the primary (uploading) author for each
paper, so instead we extracted authors from the \latex\ source. 
There are many different ways to list the authors of paper, depending
on the style file used. 
We build scripts to cover the different examples that we observed, and
attached an author number to each paper.
Among all CS papers, the average number of authors per paper is 1.72,
while for Math papers, the average number of authors is 1.24. 
38\% of CS papers have a single author, while more than half of Math
papers are written by just one author. 

Figure \ref{fig:authorwords} shows the relationship between number of
authors and the length of their papers and comments, measured in words. 
We might expect that the length of papers should grow with the number
of authors, as each author feels that they have to contribute
something extra to the paper.  
However, we do not observe a very strong relation
(Figure~\ref{fig:authortext}): length seems to be
fairly stable. 
For comments, there does seem to be a slight growth in the amount of
comment words as the number authors rises from 1 to 2 to 3
(Figure~\ref{fig:authorcomments}). 
So while comments may be used for discussion among authors, this does
not dramatically change their length. 
The behavior seems more varied for more than 6 authors, but 
there are few papers with this many authors, so there is less support
for these observations.

\subsection{Theorems}
Mathematical knowledge is typically codified in the form of theorems. 
Indeed, Erd\"{o}s defined a mathematician as ``a device for turning coffee
into theorems''\footnote{It has also been remarked that society might
  place greater value on a machine that works in the opposite direction.}. 
There are many ways to define theorems in \latex, but for our dataset we
built scripts to extract theorems based on common patterns. 
We confirmed that theorem use is more characteristic of math: at least 71\% of
Math papers contain a theorem, while only 48\% of CS papers contain
theorems. 
However, for papers with theorems, the distribution is not so
different: CS papers have 4.85 theorems on average, while Math papers
have 5.51. 

Figure~\ref{fig:paperlen_thm} shows how the length of papers varies as
a function of the number of theorems. 
Both CS and Math seem to show a similar trend, which is quite
consistent: each theorem lengthens the paper by around 600 words. 
This seems plausible: the statement, discussion and proof of a theorem
should require some reasonable amount of additional text.

\subsection{Comparison between Math and CS}
\label{sec:comparison}
Finally, we summarize the similarity and difference between CS
papers and Math papers. 

\begin{table}[t]
\centering
\caption{Comparison between Math and CS\label{tb:mathvscs}}{\small
\rowcolors{1}{}{LightSkyBlue}
\begin{tabular}{l|l|l}
  Trends & CS & Math \\  
  \hline
  submitted more than one files  & 84\% & 34\% \\
  papers with no comments   & 4.7\% & 9.6\% \\
  average number of words in comments & 772 & 395 \\
  number of pages most papers have  & 6 & 10 \\
  papers without any packages & 13\% & 25\% \\
  average packages included in one paper  & 6.7 & 5 \\
  papers using \textsf{\textbackslash newcommand} &64\% & 66\% \\
  average \textsf{\textbackslash newcommand} usages per paper &39.7 &36.1 \\
  papers having theorems  & 48\% & 71\% \\
\end{tabular}
}
\end{table}

\paragraph{Non-textual features.}
Table~\ref{tb:mathvscs} lists the key statistics that we have studied,
and presents the values for each subject. 
While some features, such as the use of theorems and use of multiple
files, are quite distinctive between the two areas, other
characteristics, such as use of \textsf{\textbackslash newcommand} are
quite similar.  

We performed a test of the predictiveness of these features, and built
a classifier that would try to predict whether a paper belonged to
Math or CS from these features alone. 
Using a logistic regression classifier, we were able to label
 81.9\% test instances correctly. 
Given such a small number of features, it is perhaps surprising that
the result is so accurate. 
Examining the parameters learned for the classifier, we saw that a
lot of weight is placed on the features 
``new commands'' and ``number of theorems''
to predict a Math paper. 
Although the likelihood of using multiple files 
is very different for Math and CS papers 
it is not significant in the classifier. 
Possibly this is because, while this feature is almost always 1 for
CS, it is more uniformly split for Math papers. 

While this showed that such features are very predictive for
different subjects, the observation does not extend to sub-categories within
areas: 
a classifier to predict which
papers were in the category cs.AI (artificial intelligence) using
the same set of features achieved only 57.4\% accuracy.

\paragraph{Textual features.}
We compared the content words of the Math and CS papers to
understand the key vocabularly difference between the two subjects. 
The ten most discriminative words for CS compared to Math are:

\centerline{\em algorithm, time, figure, data, number, 
state, model, information, probability, problem}

while the 
top-10 most discriminative words for Math compared to CS are:

\centerline{\em equation, let, alpha, lambda, infty, 
omega, frac, gamma, mathbb, map.}

While these terms should be intelligible to researchers in either
field, it is clear that notions such as ``data'' and
``information'', techniques such as ``probability'' and ``algorithm''
and concerns such as ``time'' are central to computer science.
Meanwhile, the words that define Math are mostly symbolic: ``alpha'',
``lambda'', ``gamma'', ``omega'', ``infty''; or for formatting in \latex,
like ``frac'' and ``mathbb''. 
Although, perhaps the best separation between the two fields comes
from looking at just the most discriminating word for each: 
for Math this word is ``equation'', while for CS it is ``algorithm''. 
This seems to tally with importance of the {\em algorithm} package for
CS noted in Section~\ref{sec:packages}. 
Note that the more obvious words `computer' and `mathematics' do not
appear in either top-10 (or, indeed, in the top-100). 

\section{Concluding Discussions}
\label{sec:concs}
\subsection{Related Work}

There has been much detailed study of individual scientists and
small groups; indeed, the area of History and Philosophy of Science is
based around this methodology. 
Yet, there has been limited large-scale study of the process of scientific
communication. 
Primarily, this is due to the lack of available data in a format
suitable for collation and analysis. 
Just as the growth of online social networks led to a revolution in
sociology and social network analysis, so we might anticipate greater
availability of scientific writing in accessible electronic form could
lead to renewed interest in this area. 

As mentioned in the introduction, bibliometrics and particularly
citation analysis has studied in great detail how scientific papers
reference each other~\cite{Atkins:Cronin:00,DeBellis:09,Moed:11}. 
Despite the size and significance of the 
\arxiv, there has been limited prior study of this resource. 
For example, in 2003 the KDD conference on data mining made available
29,000 papers from the high-energy physics domain, and invited
researchers to perform analysis on them\footnote{See 
\url{http://www.cs.cornell.edu/projects/kddcup/datasets.html}}.
However, the analysis published on this data concentrated almost
exclusively on the bibliographic content of the papers, and
identifying the link structure between papers, rather than any aspect
of the writing style or content. 

\subsection{Comments on Scienceography}

At one end of the process, there are many anecdotes about how
discovery and breakthroughs occur in Science; at the other end,  
bibliometrics concerns itself with the after-effects of scientific
publication, of citation analysis. Between these ends of discovery and
dissemination of a publication, we have relatively little insight into
how the writing of science is performed and how the description of
science  is compiled.  
We have identified the study of this part of the
scientific method as a topic of interest and coined the term {\em
  scienceography} (meaning ``the writing of science'') to frame the
area.  

In the past, there has been very little visibility into this aspect,
but we have made a case that with the availability of \latex\ source
in \arxiv\ together with the timestamp,   we have a data source where
certain basic aspects of scienceography can be studied.  
There is much more to be done expanding the empirical studies in 
Scienceography, as well as identifying the basic principles  and developing a theory of Scienceography. 

\medskip
\noindent
{\bf Expanding Empirical Studies.} We suggest two directions. 
Getting access to version control information used in writing papers can
provide more insights into how research papers are
composed\footnote{Similar analyses have been performed on open source
  code, such as the Linux kernel, \url{http://www.vidarholen.net/contents/wordcount/}}.
 For example, it is common to expect that papers are produced in various
sections (perhaps by different authors) and then combined with various
``passes'' by different authors.  
Does the data validate this model?
There are portions that are written and then removed from final
publication. Can we examine the intermediate forms of a research paper
and its evolution over time?  
At a more detailed level, can we quantify the ``effort'' (in terms of
time and author hours) needed for producing portions of the paper, and
indeed predict time needed from current state to the final state,
given the portions that need to be generated (and predict the probability of making a deadline for 
a conference or a grant proposal)?  
Going beyond research papers, we can consider research
presentations. 
There is data on the web which not only consists of the powerpoint
slides, but also ``comments'' in the form of author notes for each
slide which are not visible to the audience during the presentation. 
What insights can these provide into delivery of research
results by speakers? 

\medskip
\noindent
{\bf Building a Theory of Scienceography.}  
There are basic questions about models at the macro level of
communities as well as at the micro level of individuals and
individual papers. 
For example, at the macro level, 
can we develop models for how writing styles and norms (say use of
packages, naming methods for theorems or figures, and others) migrate
from community to community?   
Can we model the time dependence of how research progresses (as seen
by uploaded publications) over time in different communities? 
At the microlevel, 
are there models of social interactions of authors that can predict
the salient---scienceographic---features of a paper?   
\qed

\smallskip
Such empirical and theoretical study of Scienceography 
has the potential to bring fundamental new understandings of Science
and research. 

\section{Acknowledgement}
\thanks{ This material is based upon work supported by the National Science Foundation Under Grant No. 0916782}
{\small
\bibliographystyle{abbrv}
\bibliography{cacm}
}

\end{document}